\documentclass[usenatbib]{article}

\usepackage{amsmath, amssymb, graphicx, natbib}
\usepackage{bm}
\usepackage{enumerate}
\usepackage[left=1in, right=1in]{geometry}

\def\ltsima{$\; \buildrel < \over \sim \;$}
\def\simlt{\lower.5ex\hbox{\ltsima}}
\def\gtsima{$\; \buildrel > \over \sim \;$}
\def\simgt{\lower.5ex\hbox{\gtsima}}

\def\[{\begin{equation}}
\def\]{\end{equation}}

\def\m@th{\mathsurround=0pt }
\def\eqalign#1{\null\,\vcenter{\openup1\jot \m@th
  \ialign{\strut\hfil$\displaystyle{##}$&$\displaystyle{{}##}$\hfil
  \crcr#1\crcr}}\,}

\newcommand{\ud}{\mathrm{d}}

\newcommand{\po}{p'} 
\newcommand{\te}{t_e}
\newcommand{\thab}{t_h}
\newcommand{\tf}{\tau}
\newcommand{\tdata}{\tf_\oplus}
\newcommand{\hab}{h}
\newcommand{\tEvolve}{T_E}

\newcommand{\muh}{\mu_h}
\newcommand{\sigh}{\sigma^2_h}
\newcommand{\mue}{\mu_e}
\newcommand{\sige}{\sigma^2_e}

\begin{document}
\title{The longevity of habitable planets and \\ the development of intelligent life}
\author{Fergus Simpson$^{1}$\thanks{Email: fergus2@icc.ub.edu}}
\date{\today}
\maketitle
\begin{small}
\noindent $^{1}$ICC, University of Barcelona, Marti i Franques 1, 08028, Barcelona, Spain.
\end{small}
\maketitle

\begin{abstract}
Why did the emergence of our species require a timescale similar to the entire habitable period of our planet?  Our late appearance has previously been interpreted by \citet{2008Carter} as evidence that observers typically require a very long development time, implying that intelligent life is a rare occurrence.  Here we present an alternative explanation, which simply asserts that many planets possess brief periods of habitability. We also propose that the rate-limiting step for the formation of observers is the enlargement of species from an initially microbial state. In this scenario the development of intelligent life is a slow but almost inevitable process, greatly enhancing the prospects of future SETI experiments such as the Breakthrough Listen project. 
    

\end{abstract}
\section{Introduction}

The first sentient life on Earth appeared after approximately four billion years, while the maximum time available (defined by our planet's habitable period) may be less than six billion.  Why should these seemingly disconnected timescales, of biological and astrophysical origins, be the same order of magnitude?  What does this tell us about the timescales on other planets? To tackle these questions, we first assess the feasible range of habitable periods and evolutionary timescales, before quantifying the selection bias which afflicts our observed values.

For simplicity we follow the conventional prescription of defining a planet's habitability based on its ability to hold liquid water on its surface. The maximum time for which a planet can remain habitable is ultimately governed by the lifespan of its host star. This upper bound varies considerably between different systems, since the lifespan of a main sequence star is strongly mass-dependent. 
Moreover, many planets fail to exploit the full main sequence lifetime, instead they become inhospitable on a relatively short timescale.  Mars is a clear example of this, as it is believed to have retained surface water for only a small fraction of the Sun's lifespan.  Habitable planets should therefore be expected to display a broad range of habitable periods. It is vital to account for this variability when interpreting the value of our own planet's habitable period, yet this appears to have been overlooked in previous studies.

Little is known about the processes which regulate the emergence of complex life. \citet{carter1983anthropic} proposed that a sequence of stochastic obstacles must be overcome before intelligent life can form. To be classified as a ``hard step", the mean completion timescale must greatly exceed the planet's habitable timescale. If only a single hard step exists, observers are equally likely to emerge at any given time within a planet's habitable period. As further steps are added to the sequence, the typical emergence time is drawn towards the end of the habitable period. This has been proposed as a possible explanation for our existence towards the latter stages of the Earth's habitable period \citep{2008Carter}. 

We define our key parameters and present the available data in \S \ref{sec:data}. Then in \S \ref{sec:selection} we formalise the relationship between the true distribution of evolutionary timescales, and what we should expect to observe, followed by a worked example in \S \ref{sec:emergence}.  In \S \ref{sec:hardstep} we take a first look at the hard step model from a Bayesian perspective, in order to assess the number of steps favoured by the data.  In \S \ref{sec:easystroll} we present a model in which advanced life consistently requires several billion years to develop, due to the time taken for sufficiently large organisms to evolve from a primordial state. Finally, the implications of these results are discussed in \S \ref{sec:conclusions} .  

\section{Observables}  \label{sec:data}

We begin by defining two quantities: $\thab$ is the total duration of a planet's habitable era, and $t_e$ is the time it would take for observers to emerge on a habitable planet given an indefinite amount of time to do so.  Our planet has been capable of supporting life for approximately $4$ Gyr, but it is unclear when it will become inhospitable. Climate models typically project catastrophic temperature rises in approximately one billion years time (excluding any anthropogenic influence). Throughout this work we shall simply assume that our two observed values are given by $\te \simeq 4$ Gyr and $\thab \simeq 5$ Gyr. As we are less concerned with absolute values, this data can be more concisely expressed as the fractional emergence time $\tf \equiv \te /\thab$. Our observed value is therefore  $\tdata \simeq 0.8$. 

\section{The Biased Observation of  \\  Habitable and Evolutionary Timescales}  \label{sec:selection} 

The ensemble of habitable and evolutionary timescales, $t_h$ and $t_e$, form probability distributions denoted by $p(t_h)$ and $p(t_e)$. The potential for correlations to exist between $t_h$ and $t_e$ has been discussed previously by \citet{1999Livio}, but here we shall treat them as independent variables. For simplicity we also assume that the mean number of observers does not strongly depend on these variables. 

The \emph{observed} duration of a host planet's habitable era, $\po(t_h)$, is related to the true distribution as follows
\[ \label{eq:observed_habitable}
\po(t_h) \propto  p(t_h) \int_0^{t_h} p(t_e)   \ud t_e \, . 
\]
Unsurprisingly, longer habitable eras are more likely to be observed than short ones. One could never observe a habitable period that was briefer than the fastest evolutionary timescale. The habitable era of an observer's host planet is therefore a biased estimator for the typical duration of other habitable eras (unless the functions are orthogonal, $\int p(t_e)  p(t_h) \ud t  = 0$).
Similarly the probability distribution for the \emph{observed} emergence time $\po(t_e)$ is
\[\label{eq:observed_evolution}
\po(t_e) \propto  p(t_e)  \int_{t_e}^\infty p(t_h)   \ud t_h \, ,
\]
illustrating that fast evolutionary paths are more likely to succeed, and thus be observed, than slow ones.

Since the observed values of $t_h$ and $t_e$ are similar,  $\te = 0.8 \thab$, this implies that $\po(t_h)$ and $\po(t_e)$ are likely to be strongly localised functions, in that they possess similar means and small variances. This would ensure that a typical population of observers will appear after a significant fraction of the planet's habitable era.

The true probability distributions $p(t_h)$ and $p(t_e)$ could involve timescales which differ by many orders of magnitude. We shall therefore refer to the means and variances of their logarithmic distributions, denoted by $\muh$ and $\sigh$ respectively.  In general there are three scenarios which generate narrow distributions in $\po(t_h)$ and $\po(t_e)$:

\begin{enumerate}[(a)]
\item{There exists a broad range of evolutionary timescales, whose probability distribution  $p(\log t_e)$ rises rapidly prior to the timescale of the typical habitable period.} \item{There exists a broad range of habitable periods, whose probability distribution $p(\log t_h)$ falls rapidly beyond the typical evolutionary timescale. }
\item{The two distributions are, by coincidence, both narrow ($\sigh \lesssim 1$; $ \sige \lesssim 1$) and possess similar characteristic timescales ($\muh \simeq \mue$). }
\end{enumerate}

\noindent Hypothesis (c) demands a high degree of fine-tuning, so we shall not consider it further. The remaining two possibilities suggest that at least one of the following statements is true: either mankind emerged unusually quickly ($\mue \gg 4$ Gyr and $\sige \gg 1$), or the Earth has an unusually long habitable period compared to most other habitable planets ($\muh \ll 5$  Gyr  and $\sigh \gg 1$). 
 
The degree of overlap between $p(t_h)$ and $p(t_e)$ determines the fraction of habitable planets which at some point harbour observers, $f_o$,

\[ \label{eq:frac_life}
\eqalign{
f_o &=  p(t_e < t_h) \cr
      &=   \int_0^\infty  p(t_h)   \int_{0}^{t_h} p(t_e)  \ud t_e   \ud t_h \, .
}
\]
Given that $\muh \ll \mue$, this ensures $f_o \ll 1$, meaning that the emergence of observers is a rare event among habitable planets. This conclusion holds for both hypotheses (a) and (b). However they differ radically in their predictions for planets which are habitable at the \emph{present time}. Now we exclude planets which are either no longer habitable, such as Mars, or are yet to become habitable. The fraction of currently habitable planets which produce intelligent life, $f_t$, is given by

\[ \label{eq:frac_life_current}
\eqalign{
f_t &=  p(t_e < t_h | \hab) \cr
     &=  p(t_e < t_h) p(\hab | t_e < t_h) \cr
     &= \frac{\int_0^\infty  t_h p(t_h)   \int_{0}^{t_h} p(t_e)  \ud t_e   \ud t_h}{\int_0^\infty  t_h p(t_h)     \ud t_h} \, .
}
\]
Now the long evolutionary timescale $\muh \ll \mue$ no longer implies a sparse habitation of habitable planets $f_t \ll 1$. In fact, provided the variance of habitable periods $\sigh$ is sufficiently large, the habitation fraction $f_t$ approaches unity. 
This result considerably boosts the prospects for SETI experiments, which would otherwise appear bleak if all habitable periods were identical ($\sigh = 0$). Note that $\sigh = 0$ had been implicitly assumed in previous studies. By relaxing this assumption, we are no longer drawn to the conclusion that the vast majority of habitable exoplanets are devoid of intelligent life. The alternative interpretation is that there were many other planets, such as Mars, which remained habitable for relatively brief periods. 

\section{Late Emergence}   \label{sec:emergence} 

We now briefly demonstrate how variability in the duration of habitable periods can lead to the first observers appearing in the latter half of their planet's habitable period. For our toy model we adopt $p(t_e) = \delta_K(t_e - \tEvolve)$, where $\tEvolve = 4 \mathrm{Gyr}$.  In other words, advanced life always requires 4 billion years to emerge. 
If the distribution of habitable periods can be described as a power law when $t_h > \tEvolve$ then

\[ \label{eq:habitable_decay}
p(t_h) \propto t_h^{-\alpha} \, \, \, (t_h > \tEvolve)  \, ,
\]
while its behaviour for $t_h < \tEvolve$ has no bearing on the observed distribution. 
For $\alpha>2$, the expected habitable time can then be derived from (\ref{eq:observed_habitable})

 \[
E(t_h)  = \frac{\alpha - 1}{\alpha - 2}  \tEvolve \, .
\]
Provided the power law is sufficiently steep $(\alpha>3)$ then $ E(t_h) <2 \tEvolve $, so the first observers are not expected to appear until the latter stages of the habitable era.

\section{Hard Step Model Revisited} \label{sec:hardstep}

In this section we explore the consequences of scenario (a), as defined above in \S \ref{sec:selection}. In particular, we focus on the hard step model originally proposed by \citet{carter1983anthropic}.  How many unlikely hurdles must be overcome before an intelligent species can emerge? This number has important implications for the abundance of life in the Universe.  If there is just a single hard step prohibiting the formation of intelligent life, then there are grounds to believe that primitive life arises frequently. The relatively rapid\footnote{The oldest confirmed microfossils date back approximately $2.7$ Gyr \citep{archer2006coupled}. A number of candidates exist which are older than $3$ Gyr \citep{schopf1987early}, however considerable doubt has been cast on these claims, since their morphologies may simply be mineral artefacts \citep{brasier2015changing, wacey20153}.} emergence of life on Earth has been interpreted as evidence that abiogenesis is a relatively straightforward process \citep{2002Lineweaver}. However, if abiogenesis was the first of several hard steps, then even primitive life is a rare occurrence \citep{hanson1998must}. The stark contrast between the predictions of the single-step and multiple-step models motivates us to utilise model selection.  The purpose of model selection is to identify which model offers the best fit to the data, while accounting for the complexity or predictiveness of the model. 

\begin{figure}
\begin{center}
\includegraphics[width=80mm]{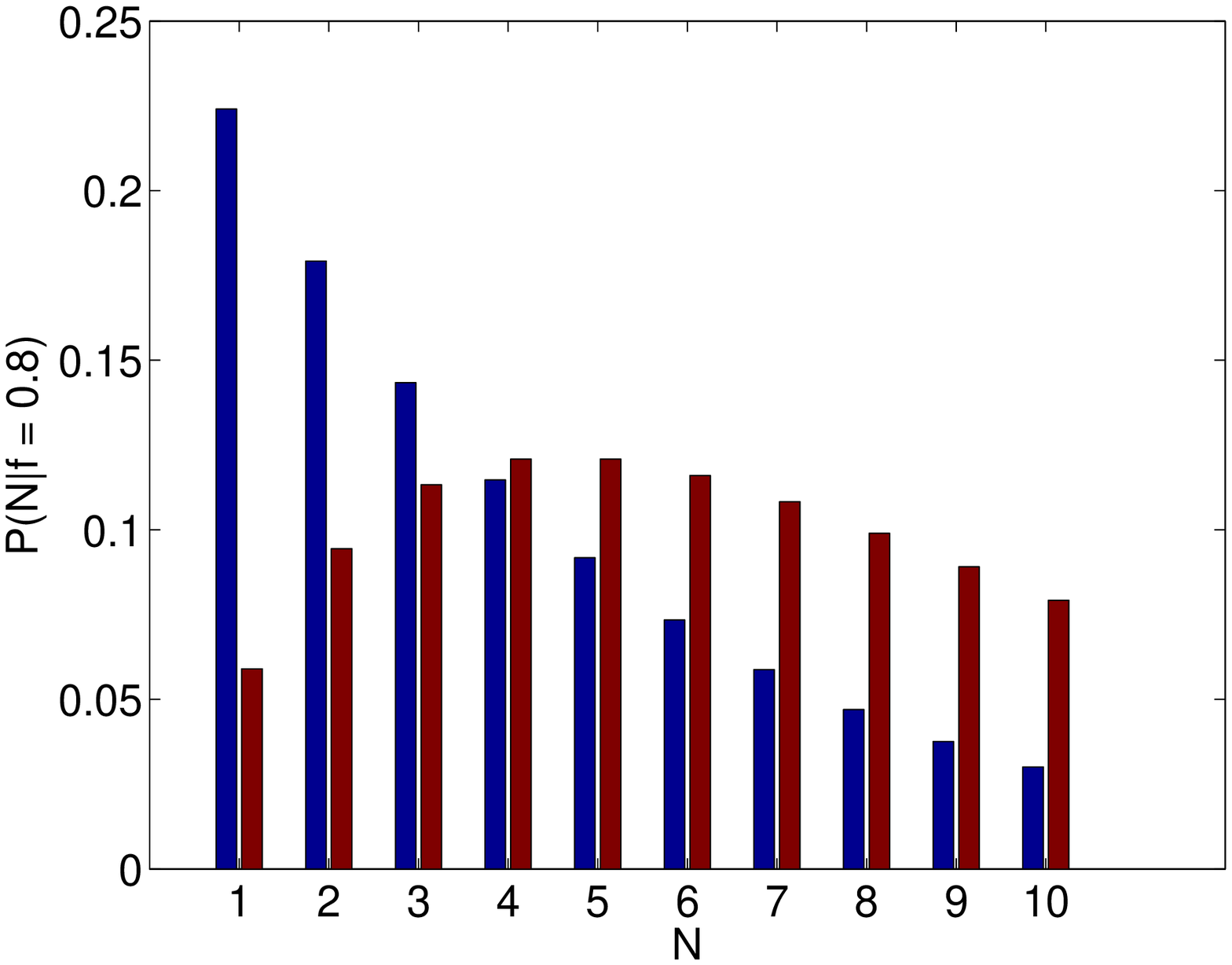}
\caption{In the context of the hard step model, the posterior probability of $N$ hard steps based on the observation of a single success arising at $80\%$ of the maximum time available. The left hand set of bars are derived from an uninformative prior $p(N) \propto 1/N$, while adopting a flat prior yields the second set of bars.
\label{fig:hardstep}}
\end{center}
\end{figure}

For $N$ hard steps, the probability that success occurs at a fractional time $\tf$ is given by \citep{2008Carter}
\[
p(\tf | N)  = \frac{N! \tf^{r-1} (\tf - 1)^{N-r}}{(r-1)!  (N - r)!} \, .
\]
If N=1 this is constant, implying that observers are equally likely to emerge at any given time, while for $N>1$ the function peaks at $\tf=1$. 
To compare the single step model $(N=1)$ with the multi-step case, where $N$ is a free parameter, we evaluate the Akaike Information Criterion (AIC), defined as 
\[
\mathrm{AIC} = 2k - 2 \log (\mathcal{L}) \, ,
\]
where $\mathcal{L}$ is the maximum likelihood and $k$ is the number of free parameters \citep{akaike1974new}.  For the single step model we find $\mathrm{AIC}_s=5.66$, while the multi-step model yields $\mathrm{AIC}_m= 6.23$, with the maximum likelihood achieved with $N=5$. The relative likelihood is given by $\mathrm{exp}((\mathrm{AIC}_s - \mathrm{AIC}_m)/2) = 0.75$.   The multi-step model is therefore weakly disfavoured relative to the simpler single-step case. If we were to adopt the Bayesian Information Criterion (BIC, \citealt{schwarz1978}), the multi-step model would be more heavily penalised. Therefore, even within the context of the hard step model,  there is no evidence to suggest that abiogenesis is a slow process, or a rare event. 

Figure \ref{fig:hardstep} provides an illustration of why there was insufficient evidence to invoke a new parameter in the model to fit the data. The two sets of bars displays the posterior probability $p(N | \tau = 0.8)$  for two different choices of prior $p(N)$. The posterior is given by Bayes' theorem
 \[ \label{eq:bayes}
 p(N | \tf) = \frac{p(\tf | N) p(N)}{p(\tf)} \, .
 \]
The left set of bars represent the case $p(N) \propto 1/N$, the conventional choice for an uninformative prior, while the second set of bars corresponds to a uniform prior. In each case, it is evident that $N=1$ is not significantly disfavoured. 

\section{Easy Stroll Model}  \label{sec:easystroll} 

The formation of the Earth did not require billions of years because it was an improbable event - many other planets formed on a similar timescale -  it required billions of years because it involved fundamentally slow processes. These include the collapse of cosmic structure, the lifespan of the first stars, and the growth of planetesimals. Similarly, the development time for mankind may have been limited by a slow process rather than a difficult one. This corresponds to scenario (b) (see \S \ref{sec:selection}) where habitable periods vary by several orders of magnitude, while evolutionary timescales are more consistent. 
 
The evolutionary trajectory from a primordial microbe to our macroscopic form demanded a change in organism size of over sixteen orders of magnitude. How rapidly could we realistically expect this transformation to take place? One of the fastest cases of gigantification to be well-documented is the trajectory followed by cetacea. Over the past $50$ million years the blue whale has maintained a remarkable growth rate, increasing in body mass by three orders of magnitude \citep{montgomery2013evolutionary}.  Even if we use this extraordinary individual example as a benchmark, at this rate we would still require over $250$ million years to rise sixteen orders of magnitude in body mass. The mean enlargement rate of the largest organism is likely to be substantially slower than it was for this particular mammal over this relatively brief timespan. It therefore appears feasible that the evolution of sufficiently large human precursors invariably takes several billion years, thereby acting as our rate-limiting step, as opposed to other obstacles such as encephalisation.

The size requirement is not a unique problem for our species, but presents a more general problem for all observers. To be classified as an observer, an organism must exceed some minimum degree of complexity, and with that requirement comes a minimum brain size. Even if we conservatively set a minimum observer body mass of $1$kg - less than the mass of the human brain -  this still leaves a formidable 15 orders of magnitude to climb. Further justification for this mass limit comes from an independent calculation, which exploits the relationship between body size and population density.  \citet{2015SimpsonAliens} estimated the minimum mass of a given species of observer to be $m>25$kg (95\% confidence limit).

\begin{figure}
\begin{center}
\includegraphics[width=80mm]{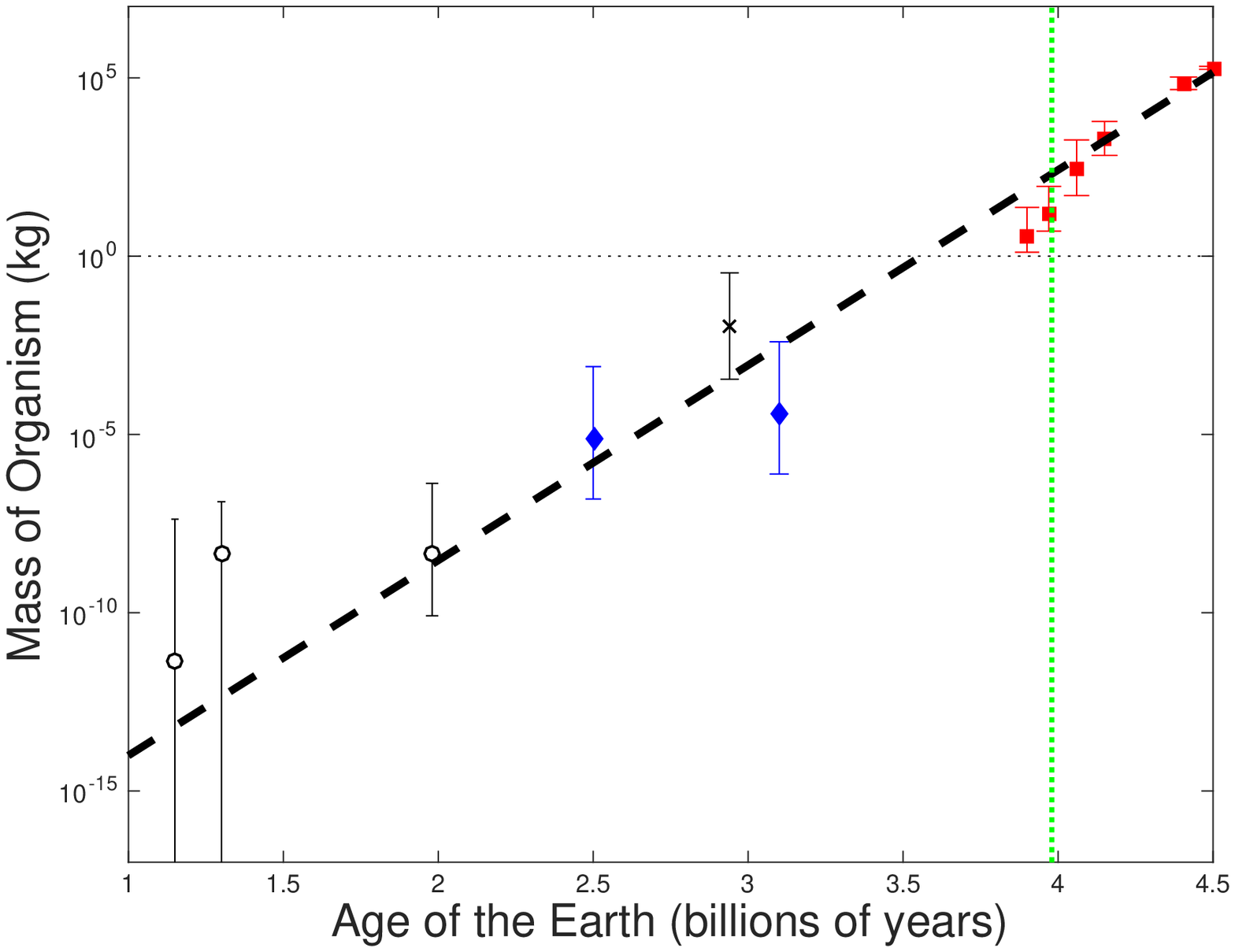}
\caption{The exponential enlargement of species over the past four billion years. The three hollow black circles relate to the organic-walled microfossils presented in \citet{wacey2011microfossils}, \citet{javaux2010organic}, and \citet{czaja2016} respectively. The upper error bars represent our highly incomplete state of knowledge for organisms at that time, while the lower error bars reflect the possibility that the microfossils comprise colonies of smaller organisms.  The two blue diamonds correspond to specimens of Grypania, thought to be a eukaryotic alga, dated at 1.4 Gyr and 2.1 Gyr  \citep{runnegar1991precambrian, han1992megascopic}.  
The six red squares correspond to the largest animal species of their time, from left to right: Anomalocaris; Dickinsonia Rex; Cameroceras; Rhizodus; Argentinasaurus; and the blue whale.
The diagonal dashed line ilustrates the trajectory expected from the simplest diffusion model, as specified by equation (\ref{eq:growth}). The vertical dashed line marks the onset of the Cambrian explosion, associated with the emergence of numerous macroscopic species.
\label{fig:slowstep}}
\end{center}
\end{figure}

Figure \ref{fig:slowstep} maps the enlargement of some of the Earth's largest organisms across a logarithmic mass scale. The plotted points reflect the largest known species of their time, which can be well determined at late times but our confidence prior to the onset of the Cambrian explosion (vertical dashed line) is considerably weaker. The earliest evidence for living organisms  comes from  \citet{wacey2011microfossils}, \citet{javaux2010organic}, and \citet{czaja2016}. The earliest recorded macroscopic species is Grypania  \citep{runnegar1991precambrian, han1992megascopic}, but again we cannot be certain either that we have truly found the largest organism of its time, or if it consists of a colony of smaller organisms. The tail of the mass distribution is highly extended, meaning that even a small degree of incompleteness can lead us to significantly underestimate the size of the largest species.

At late times we focus on the largest species within the animal kingdom, where fossil records show a progressive enlargement, from Anomalocaris, Dickinsonia Rex, Cameroceras, Rhizodus and Argentinasaurus,  while the present day is represented by the blue whale. Note that the plant kingdom followed a very similar growth pattern, as illustrated in Figure 1 of \citet{payne2009two}. 

The spread of the biosphere's mass function (which we define as the abundance of biomass as a function of the organism's body mass) appears to have been broadly consistent with an exponential growth law. This behaviour is to be expected from a multiplicative diffusion process, a model which has been substantiated by the diversification of mammalian species \citep{clauset2009evolutionary} and the observed distributions  of inter-species and intra-species body masses \citep{clauset2008evolution}.  

 \citet{payne2009two} concluded that the fossil records are highly inconsistent with a simple diffusion model. This was partially based on the assertion that a random walk progresses in proportion to the square root of time. However in a random walk it is the \emph{median} path, not the \emph{maximum} path, which propagates as the square root of the number of steps taken. The maximum path evolves approximately linearly with time, as illustrated by the solid line in Figure \ref{fig:slowstep}. If the maximum organism mass is denoted by $m$ then its evolution with time may be parameterised as
 
 \[ \label{eq:growth} 
m = m_0 e^{t/\tau} \, .
\]
where $m_0$ is the initial organism size at time $t=0$,  $\tau = \ln(2) t_{2}$ is the growth constant, and $t_{2}$ is the timescale for the maximum organism size to double.  For Figure \ref{fig:slowstep} we adopt a doubling time of $t_{2} = 55$ million years, as shown by the dashed line. 

Now we consider the error bars attributed to the data points. For larger and more recent species the detection efficiency is expected to be high. However it seems extremely unlikely that the detection efficiency remains as high for the older microscopic organisms. For example, the fossils of the largest known dinosaur species were found not by paleontologists but a local farm worker. Such serendipitous finds clearly cannot occur with microfossils. Furthermore it can be difficult to distinguish a single large microorganism from a colony of smaller microorganisms. We therefore limit ourselves to a few early cases prokayotes and eukarotes, and attribute large uncertainties to the upper bound of their masses. Note that unlike \citet{payne2009two}, we do not include the $3.5$ Gyr Primaevifilum \citep{schopf1987early}, since recent studies have found these morphologies to be consistent with mineral artefacts \citep{brasier2015changing, wacey20153}. Instead we consider the $3.2 - 3.4$ Gyr cases documented by \citet{javaux2010organic}, \citet{wacey2011microfossils}, \citet{sugitani2015early}, and the $2.5$ Gyr samples from \citet{czaja2016}. These studies were all published in the last few years, indicating the rapid development of the field. 

Of the macroscopic species shown, only Cameroceras has a poorly determined mass. The size of this mollusc is extremely challenging to estimate. For the largest candidates, only small shell fragments have been found, leading to conflicting extrapolations of its full size. A further complication is that around $90\%$ of its volume consists of a hollow shell, rather than biological tissue. 

There were undoubtedly a number of key events along this trajectory such as the first cell nucleus, the Great Oxygenation Event, and various mass extinctions \citep{szathmary1997replicators, szathmary2000major}. However it is often the case that simple  scaling behaviours emerge from highly complex systems. For example, over the past fifty years the development of semiconductor technologies has adhered to the smooth progression of Moore's Law, despite relying upon a series of individual breakthroughs such as the invention of photolithography. 


 



Following the acceptance of this paper, \citet{zhu2016decimetre} announced the discovery of large multicellular eukaryotes, from a rock formation identified to be 1.56 billion years old. The largest of these organisms was found to have dimensions of approximately $28.6$cm by $7.6$cm. Their thickness is highly uncertain, but is estimated at approximately 1mm.  This new finding has been added to Figure 2, with the datapoint marked as a cross, and appears consistent with our theoretical model, simply described by an ensemble of random walks.

\section{Conclusions} \label{sec:conclusions}



Few hypotheses can naturally explain our appearance towards the latter stages of the Earth's habitable period. One possibility is that our evolution was fortuitously rapid, which in turn implies that a large fraction of habitable planets never produce intelligent life. In this work we have highlighted an alternative explanation, in which many planets possess relatively brief habitable periods. This lifts the previous restrictions on the prevalence of intelligent life, which is encouraging news for future SETI experiments. 

When interpreting the duration of the Earth's habitable period, we must account for the strong selection effects which are at work. Some planets will sustain habitable conditions for longer than others, and we are likely to find ourselves on a planet with a long habitable period. That habitable periods can vary in their duration appears to have been overlooked in previous studies. Transient periods of habitability, as exhibited by Mars, may be a much more frequent phenomenon than environments which remain continuously stable for many billions of years. Aside from atmospheric mass loss, habitable periods might be truncated by unstable climates. These instabilities could arise from positive feedback mechanisms such as the heightened albedo following the formation of ice, or a runaway greenhouse effect following an increase in surface temperatures. 

In this work we have assumed that the mean population of observers is uncorrelated with the length of the habitable period.
Complications arise if the mean population of observers is correlated with the length of the habitable period. The probability distributions defined in (\ref{eq:observed_habitable}) and (\ref{eq:observed_evolution})  then ought to include a weighting function  $\bar{n}(t_h, t_e)$. The strength of population selection effects can be substantial \citep{2015SimpsonAliens}. As one example of how this could arise, consider the population of a hypothetical civilisation hosted by an M dwarf. Since the stellar irradiance is predominantly in the infra-red, only a small proportion reaches the surface. With less energy available, the planet can only support a small number of observers. This is further exacerbated if tidal locking renders most of the planet's surface uninhabitable. This kind of population dependence acts as another mechanism to bias the observed values of $t_h$ and $t_e$. 

We revisited the hard step model, finding that our appearance towards the end of our habitable period is insufficient evidence to introduce multiple steps. From a Bayesian perspective, the simpler case of a single hard step is preferred, as it avoids overfitting the data.

The timescale for the emergence of observers may be dictated by an intrinsically slow process: the enlargement of species. Given the rate at which this process is known to have occurred, there is little reason to believe that primitive life could gain fifteen orders of magnitude in mass in much less than one billion years. Planets younger than our own are less likely to host intelligent life. Stellar age could therefore be a useful indicator for prioritising targets in SETI experiments such as Breakthrough Listen - there is little hope in listening to younger star systems.

\section*{Acknowledgments}

The author would like to thank Andrew Liddle and David Wacey for helpful comments, and acknowledges support by the European Research Council under the European Community's Seventh Framework Programme FP7-IDEAS-Phys.LSS 240117. 

\bibliographystyle{mn2e} 
\bibliography{../../../HomeSpace/Routines/dis} 

\end{document}